\documentclass[12pt]{article}

\usepackage{amssymb}
\usepackage{amsmath}
\usepackage{citesort}
\usepackage{graphicx}

\makeatletter
\@addtoreset{equation}{section} 
\makeatother

\renewcommand{\theequation}{\arabic{section}.\arabic{equation}}

\newcommand{\bq}{\begin{eqnarray}}
\newcommand{\eq}{\end{eqnarray}}
\newcommand{\bqn}{\begin{eqnarray*}}
\newcommand{\eqn}{\end{eqnarray*}}
\newcommand{\ff}{{\cal F}}
\newcommand{\rr}{{\mathbf r}}
\newcommand{\kk}{{\mathbf k}}

\begin{document}
\begin{titlepage}

\title{{\bf Generating functionals, consistency, and uniqueness in
the integral equation theory of liquids}}

\author{{\bf R. Fantoni}$^1$ and {\bf G. Pastore}$^1$}
 
\maketitle

\begin{abstract}
\noindent
We discuss and illustrate through numerical examples the relations between
generating functionals, 
thermodynamic consistency (in particular the virial-free energy one), 
and uniqueness of
the solution, in the 
integral equation theory of liquids. We propose a new approach
for deriving  closures  automatically satisfying such
characteristics. Results from a first exploration of this program are presented
and discussed.
\end{abstract}

\vfill

\noindent $^1$Dipartimento di Fisica Teorica dell' Universit\`a  
and Istituto Nazionale di Fisica della Materia, Strada Costiera 11, 
34014 Trieste, Italy; e-mail: rfantoni@ts.infn.it and
pastore@ts.infn.it 

\end{titlepage}

\section[Introduction]{Introduction}
\label{sec:introduction}

Integral equation theories (IET) of the liquid state statistical 
mechanics are 
valuable tools for studying structural and thermodynamic properties 
of pairwise interacting fluid systems \cite{Hansen,Caccamo}. Many of these 
approximations to the exact relation between pair potential and pair 
correlation functions have been proposed in the last half century, 
starting from the pioneering works \cite{BornGreen,PY,HNC} to the most refined 
and modern approximations 
\cite{Rosenfeld79,Lado,Verlet,MS,Duh}
which may approach the accuracy of computer 
simulation with a negligible computational cost. 

The functional method in statistical mechanics \cite{Hansen} provides the most 
general and sound starting point to introduce IET as approximations of
the exact functional relations and it is the classical statistical 
mechanics counterpart of the quantum density functional theory.

Notwithstanding the success of present IET to describe the structure 
of simple one component systems, considerable work is still devoted to 
derive improved approximations which could accurately describe the 
thermodynamics as well. Also applications to  non 
simple or multicomponent systems are still subject of current studies. 

Actually, the description of thermodynamics is one  weak point of IET 
approaches:  reasonable and apparently harmless approximations to the 
potential-correlation relations usually result in a dramatically 
inconsistent thermodynamics where many, if not all, among the exact sum rules 
derived from statistical mechanics,  are violated.

The problem of thermodynamic inconsistency, {\sl i.e}. the inequivalence 
between different routes to thermodynamics,  actually plagues the IET 
approach to the point that the degree of inconsistency between different 
formulae for the same quantity is used as an intrinsic measurement of the 
quality of 
a closure.

In the past, some discussion of the thermodynamic consistency appeared 
in the literature. Hypernetted chain approximation (HNC)  was 
recognized as a closure directly derivable  from an approximation for the 
free energy functional \cite{Morita61} , thus exhibiting consistency
between the virial formula and the thermodynamic expression for the
pressure. However, this limited consistency is not enough to guarantee
a unique and faithful description  of the phase diagram. Apart the
problem of the remaining inconsistencies, the descriptions of the
critical points and spinodal lines are seriously inadequate.

Extensive work on HNC \cite{Belloni93,AshcroftPoll,TDG} showed that in 
place of a true spinodal line, it is more appropriate to describe the 
numerical results as due to a region in the thermodynamic plane 
where no real solution of the integral equation exists. In 
particular, Belloni\cite{Belloni93} showed that the disappearance of the 
solution 
originates from a branching point where two solutions merge, instead than from
a line of diverging compressibility. Thus, we 
have direct evidence that HNC may have multiple solutions, at least in 
part of the phase diagram.

Empirical improvements on HNC have been proposed \cite{Rosenfeld79,MS,Duh} 
providing in many cases excellent results for one-component simple 
fluids. However, although reduced, the thermodynamic inconsistency problem 
remains and the multiple solution problem is completely untouched.

In this work we start an investigation of a new approach to IET 
directly addressing the two points of uniqueness of the solution and 
thermodynamic consistency. The 
basic idea is to constrain the search for new closures within the 
class of generating functionals which are strictly convex  
free energy functionals, thus enforcing the virial-energy consistency 
as well as the uniqueness of the solution.

In particular, in the present paper we try to answer the following 
questions: i) does at least one strictly convex 
free-energy functional of the pair correlation function exist? ii) 
what is the 
nature of the resulting spinodal line (if any), iii) what is the
quality of the  
resulting thermodynamic and structural results? iv) does the 
simultaneous requirement of consistency and uniqueness automatically 
provide improved results?

As we will show, we have a positive answer for i), a thorough and 
interesting characterization for ii), some interesting indications for iii),  
and a partly negative answer for iv).

However, we can show that it is possible to exploit the control provided
by the generating functional approach to easily generate new closures and  
we feel our procedure could be the basis of a 
more systematic approach to IET.

In section \ref{sec:uniqueness}  we recall the connections between
closures, generating 
functionals, thermodynamic consistency and  uniqueness of solutions and we
illustrate them in the well known case of HNC approximation. In
section \ref{sec:extensions} we introduce two straightforward 
extensions of HNC intended to cure its problems. In Section
\ref{sec:results} numerical results are presented and discussed. In
section \ref{sec:improving} we show two possible improvements of the
closures studied. 

\section[Thermodynamic consistency and uniqueness of the solution of 
integral equations]{Thermodynamic consistency and uniqueness of the
solution of integral equations}
\label{sec:uniqueness}

Since the work by Olivares and McQuarrie \cite{Olivares76} it is 
known the general method to obtain the generating functional whose 
extremum with respect to variations of the direct ($c(r)$) or total 
($h(r)$) correlation functions results in the closure relation,
provided the Ornstein-Zernike equation is satisfied.

For example, if we have a closure of the form 
\begin{equation}
\rho^2 c(r) = \Psi\{h(r),\beta \phi(r)\}~~~,
\end{equation}
where $\phi(r)$ is the pair interaction potential and $\Psi$ is an arbitrary 
function, the functional 
\bq \nonumber
Q[h(r),\beta\phi(r)]&=&\frac{1}{2\beta\rho}\left(\int
\frac{d\kk}{(2\pi)^3}\{\rho h(k)-\ln[1-\rho h(k)]\}\right.\\
&&-\left.\int d\rr\,h(r)\int_0^1dt\,\Psi\{th(r),\beta\phi(r)\} 
+\mbox{constant}\right)~~~, 
\eq
is such that the extremum condition
\begin{equation}
\frac{\delta Q }{\delta h(r)} = 0~~~,
\end{equation}
is equivalent to
\bq
\rho^2 h(r)=\Psi\{h(r),\beta \phi(r)\}+\rho\int h(|\rr-\rr^\prime|)
\Psi\{h(r^\prime),\beta \phi(r^\prime)\}\,d\rr^\prime~~~.
\eq

Olivares and McQuarrie also showed how to find the generating functional if the
closure is expressed in the form
\begin{equation}
\rho^2 h(r) = \Psi\{c(r),\beta \phi(r)\}~~~.
\end{equation}

In appendix \ref{app:gammafunct} we discuss the extension of their
method to the case of a closure written as 
\begin{equation}
\rho^2 c(r) = \Psi\{\gamma(r),\beta \phi(r)\}~~~,
\end{equation}
where $\gamma(r)= h(r) -c(r)$ is the indirect correlation function.
Notice that most of the modern closures correspond to this last case.

The possibility of translating the original integral equation into an 
extremum problem allows to get an easy control on two important 
characteristics of the approximation: thermodynamic 
consistency between energy and virial  routes to the thermodynamics 
and  uniqueness of the solution.

Indeed, once we get the generating functional $Q$, due to the 
approximations induced by the closure, there is no guarantee that its value  
at the extremum  is  an excess free energy.
In order to be a free energy, the functional should satisfy  the  
condition 
\begin{equation} \label{consis}
\frac{\delta Q }{\delta \phi(r)} = \frac{\rho}{2} g(r)~~~,
\end{equation}
where $g(r)=h(r)+1$ is the pair distribution function.

Even if this condition is not new, and mention to it is present in 
the literature \cite{Hoye}, we discuss it in appendix \ref{app:consistency}
as well as its consequences on  the 
thermodynamic consistency between the virial pressure and the 
density derivative of the 
free energy.

Another issue where the generating functional approach is useful is the 
problem of multiple solutions of the integral equations \cite{Belloni93}.
In particular, the analysis of the convexity properties of the 
generating functional is a very powerful tool \cite{MSA,ORPA}.

Let us illustrate this techniques in the case of HNC closure.
It is well known \cite{Morita61,Olivares76} that the HNC equation with closure
\bq
c(r) = h(r) - \ln\left[g(r)e^{\beta\phi(r)}\right]~~~,
\eq
can be derived from the variational principle
\bq \label{variational}
\frac{\delta \ff[h]}{\delta h(r)}=0~~~,
\eq
where 
\bq
\ff[h]=\ff_{OZ}[h]+\ff_{HNC}[h]~~~,
\eq
with 
\bq
\left\{
\begin{array}{l}
\displaystyle\ff_{OZ}[h]=\int\frac{d\kk}{(2\pi)^3}\{\rho\hat{h}(k)-
\ln[1+\rho\hat{h}(k)]\}~~~,\\\\
\displaystyle\ff_{HNC}[h]=\rho^2\int d\rr\left\{1+g(r)\left[\ln\left(g(r)
e^{\beta\phi(r)}\right)-1\right]-h^2(r)/2\right\}~~~.
\end{array}
\right.
\eq

Let us call $\bar{h}(r)$ the extremum of $\ff$, solution of the
variational principle (\ref{variational}). It can be shown (see
appendix \ref{app:consistency}) that, within an   
additive constant, $\ff[\bar{h}]/(2\beta\rho)$ is the excess Helmholtz 
free energy per particle of the liquid. This ensure thermodynamic 
consistency between the route to the pressure going through the partial 
derivative of the free energy and the one going through the virial
theorem (see appendix \ref{app:consistency}). In addition, it allows
to get a closed 
expression for the excess chemical potential without further approximations
\cite{Kjellander,Lee92}. This feature is highly desirable for
applications of IET to 
the determination of the phase diagrams.

Moreover if we can prove that $\ff$, defined on some convex set of 
trial correlation functions $D_c$, is a strictly convex functional, 
then we know that 
if a solution to (\ref{variational}) exists, it corresponds to a 
minimum and is unique. A functional $\ff$ is strictly convex if for 
all $y(r)\in D_c$ and $y(r)\neq 0$, we have
\bq
A=\int y(r)\frac{\delta^2\ff[h]}{\delta h(r)\delta h(r^\prime)} 
y(r^\prime)\,d\rr\,d\rr^\prime>0~~~.
\eq
We calculate the second functional derivatives as follows
\bq
\left\{
\begin{array}{l}
\displaystyle\frac{\delta^2\ff_{OZ}[h]}{\delta h(r)\delta h(r^\prime)}
=\rho^2\int\frac{d\kk}{(2\pi)^3}e^{-i\kk\cdot(\rr+\rr^\prime)}\frac{1}
{[1+\rho\hat{h}(k)]^2}~~~,\\\\
\displaystyle\frac{\delta^2\ff_{HNC}[h]}{\delta h(r)\delta h(r^\prime)}
=\rho^2\delta(\rr-\rr^\prime)\left(\frac{1}{g(r)}-1\right)~~~.
\end{array}
\right.
\eq
Recalling that the static structure factor $S(k)=1+\rho\hat{h}(k)$, we 
find for $A$
\bq \label{aor2}
A/\rho^2=\int\frac{d\kk}{(2\pi)^3}\frac{\hat{y}^2(k)}{S^2(k)}+\int
d\rr\,y^2(r)\left(\frac{1}{g(r)}-1\right)~~~.
\eq
Now, the most interesting results would be to show the strict 
convexity of the HNC functional over the convex set of all the 
admissible pair correlation functions (all the  $h(r) \geq -1$ and 
properly decaying to zero at large distance.

However,  this is not the case for HNC. It has not been possible to 
show the positive definiteness of equation (\ref{aor2}) and it has
been shown \cite{Belloni93} that in some region of the thermodynamic
plane HNC does exhibit multiple solutions.

The best we can do is to obtain a more limited result.
Calling $g_1=\sup{g(r)}$ ($g_1>1$ is the height of the first peak of the 
pair distribution function) and using Parseval theorem, we find 
\bq
A/\rho^2>\int\frac{d\kk}{(2\pi)^3}\hat{y}^2(k)\left(\frac{1}{S^2(k)}
-1+\frac{1}{g_1}\right)~~~,
\eq
from which we deduce that $A>0$ on the following set of functions
\bq
D=\left\{h(r)\,|\,0<S(k)<\sqrt{g_1/(g_1-1)}~~~\forall k\right\}~~~.
\eq
We conclude that $\ff$ defined on any convex set of functions $D_c
\subset D$ is strictly convex. Near the triple point we are sure we 
are out from such set since the
first peak of the pair distribution function for the Lennard-Jones
fluid is $g_1\simeq 3$ \cite{Verlet68}, so that
$\sqrt{g_1/(g_1-1)}\simeq 1.2$. The first peak of the static structure
factor is also close to 3. Then we are not inside $D$ and the HNC
approximation may have multiple solutions \cite{Belloni93}.

Instead, if we are in the weak coupling regime, the previous 
conditions tells us that there is a range where the branch of 
solutions going to the perfect gas limit is unique and quite isolated 
from other solutions.

\section[Extensions of HNC]{Extensions of HNC}
\label{sec:extensions}

The generating functional approach can be used in a systematic way 
to look for better closures. We think that this way, we can obtain a less  
empirical search method for improving closures.

In the following we report some preliminary analysis we have done.
As a first test of our program, we have restricted our investigations to 
simple modifications of HNC 
functional. As we will discuss later, such a choice is certainly  not optimal. 
However, we can learn enough to consider the approach worthwhile of 
further investigations and we feel the results are interesting in 
order to reveal more details about the characteristics of the 
solutions of the highly non linear IET.

\subsection[The HNC/H2 approximation]{The HNC/H2 approximation}
\label{subsec:HNC/H2} 

We want to modify the HNC closure in order to have an integral equation 
with a generating functional which is strictly convex without having to 
restrict its definition domain. We choose as our modified HNC (HNC/H2) 
closure \cite{footnote}
\bq
c(r)=h(r)-\ln[g(r)]-\beta\phi(r)-\alpha h^2(r)~~~,
\eq
with $\alpha$ a parameter to be determined. The new closure generating 
functional is
\bq
\ff_{HNC/H2}[h]=\rho^2\int d\rr\left\{1+g(r)\left[\ln\left(g(r)
e^{\beta\phi(r)}\right)-1\right]-h^2(r)/2+\alpha h^3(r)/3\right\}~.
\eq
Its second functional derivative with respect to $h$ is
\bq
\frac{\delta^2\ff_{HNC/H2}[h]}{\delta h(r)\delta h(r^\prime)}=
\rho^2\delta(\rr-\rr^\prime)\left[\frac{1}{g(r)}-1+2\alpha h(r)\right]~~~.
\eq
Recalling that $h=g-1$ and $g(r)>0$ for all $r$, we see that for 
$\alpha=1/2$ 
\bq
\frac{1}{g}-1+2\alpha h=\frac{(1-g)^2}{g}\ge 0~~~\forall g~~~.
\eq
Then $\ff_{HNC/H2}$ is a convex functional and since $\ff_{OZ}$ is 
unchanged and strictly convex (see appendix \ref{app:convexity}), their sum, 
the generating functional of the integral equation, is strictly convex. 

Moreover $\{\ff_{OZ}[\bar{h}]+\ff_{HNC/H2}[\bar{h}]\}/(2\beta\rho)$
continues to be the excess Helmholtz free energy per particle of the
liquid since equation (\ref{consis}) holds (see appendix \ref{app:consistency}). 

We have then an integral equation which is both thermodynamically 
consistent (the pressure calculated from the virial theorem coincides
with that one calculated from the Helmholtz free energy) and with a
solution which, when it exists, is unique.

\subsection[The HNC/H3 approximation]{The HNC/H3 approximation}
\label{subsec:HNC/H3}

In the same spirit as in subsection \ref{subsec:HNC/H2} we can try to
add a term $h^3$ in the HNC/H2 closure
\bq
c(r)=h(r)-\ln[g(r)]-\beta\phi(r)-\alpha h^2(r)-\gamma h^3(r)~~~,
\eq
with $\alpha$ and $\gamma$ parameters to be determined. We call
this approximation HNC/H3. The closure generating functional is
\bq \nonumber
\ff_{HNC/H3}[h]=\rho^2\int d\rr&&\hspace{-.5cm}\left\{1+g(r)\left[
\ln\left(g(r)e^{\beta\phi(r)}\right)-1\right]-h^2(r)/2\right.\\
&&\left.+\alpha h^3(r)/3+\gamma h^4(r)/4\right\}~~~.
\eq
Its second functional derivative with respect to $h$ is
\bq \nonumber
\frac{\delta^2\ff_{HNC/H3}[h]}{\delta h(r)\delta h(r^\prime)}&=&
\rho^2\delta(\rr-\rr^\prime)\left[\frac{1}{g(r)}-1+2\alpha h(r)
+3\gamma h^2(r)\right]\\ 
&=&\rho^2\delta(\rr-\rr^\prime)\frac{1-g(r)}{g(r)}\{1-2\alpha g(r)
+3\gamma g(r)[1-g(r)]\}~~~.
\eq
In order to have the right hand side of this expression positive for 
$g>0$ the only choice we have is to set $\alpha=1/2$. In this way
\bq
(1-g)[1-2\alpha g+3\gamma g(1-g)]=(1-g)^2(1+3\gamma g)~~~,
\eq
and we see that $\ff_{HNC/H3}$ is a convex functional if we
additionally choose $\gamma> -1/[3\sup g(r)]$. 

Once again
$\{\ff_{OZ}[\bar{h}]+\ff_{HNC/H3}[\bar{h}]\}/(2\beta\rho)$ is the
excess Helmholtz free energy per particle of the liquid and the
thermodynamic consistency virial-free energy is ensured.

\section[Numerical results]{Numerical results}
\label{sec:results}

To solve numerically the OZ plus closure system of nonlinear
equations we used Zerah' s algorithm \cite{Zerah85} and
Fourier transforms were done using  fast Fourier transform. 
In the code we always work with adimensional thermodynamic variables 
$T^*=1/(\beta\epsilon)$, $\rho^*=\rho\sigma^3$, and
$P^*=P\sigma^3/\epsilon$, where $\sigma$ and $\epsilon$ are
the characteristic length and characteristic energy of the system
respectively. We always used 1024 grid points and a step
size $\Delta r=0.025 \sigma$.

The thermodynamic quantities were calculated according to the statistical
mechanics formulae for: the excess internal energy
per particle
\bq
U^{exc}/N=2\pi\rho\int_0^\infty \phi(r)g(r)r^2\,dr~~~,
\eq
the excess virial pressure
\bq
\beta P^{v}/\rho-1=-\frac{2}{3}\pi\beta\rho\int_0^\infty
\frac{d\phi(r)}{dr}g(r)r^3\,dr~~~,
\eq
the bulk modulus calculated from the compressibility equation
\bq
B_c=\frac{\beta}{\rho\chi_T}=\frac{1}{S(k=0)}~~~,
\eq
where $\chi_T$ is the isothermal compressibility, and the bulk modulus
calculated from the virial equation
\bq \nonumber
B_p=\beta\frac{\partial P^{v}}{\partial\rho}~~~.
\eq
For the calculation of $B_p$ once $g(r)$ and $c(r)$ had been
calculated, Lado' s scheme for Fourier transforms \cite{Lado71} was
used to determine $\partial\hat{g}(k)/\partial\rho$. Even if slow,
this allows us to explicitly calculate and later invert the
coefficients matrix of the linear system of equations which enters the
calculation of $\partial\hat{g}(k)/\partial\rho$.

\subsection[Inverse power potentials]{Inverse power potentials}

The general form of the inverse power potential is
\bq
\phi(r)=\epsilon\left(\frac{\sigma}{r}\right)^n~~~,
\eq
where $3<n<\infty$. For this class of fluids the thermodynamics depends only
from the dimensionless coupling 
parameter 
\bq
z=(\rho\sigma^3/\sqrt{2})(\beta\epsilon)^{3/n}~~~.
\eq

We performed our calculations on the $n=12, 6, 4$ fluids at the 
freezing point. We compared three kind of closures: the 
one of Rogers and Young \cite{Rogers84} (RY) with thermodynamic
consistency virial-compressibility and known to be very close to the simulation
results, the hypernetted-chain (HNC)
closure, and the HNC/H2 described in subsection \ref{subsec:HNC/H2}. 
In each case we compared our data with the Monte Carlo (MC) results 
of Hansen and Schiff \cite{Hansen73}.

\subsubsection[The inverse 12th power potential]{The inverse 12th power potential}

The freezing point for this fluid is at $z=0.813$. The RY $\alpha$
parameter to achieve thermodynamic consistency at this value of $z$ is
$0.603$.  

In table \ref{tab:12} we compare various thermodynamic quantities
(the excess internal energy per particle, the excess virial pressure,
the bulk moduli) obtained from the RY, the HNC, and the HNC/H2 closures. 
In the MC calculation the excess virial pressure
is 18.7 and the bulk modulus 72.7.

\begin{figure}[hbt]
\centerline{\includegraphics[width=10cm]{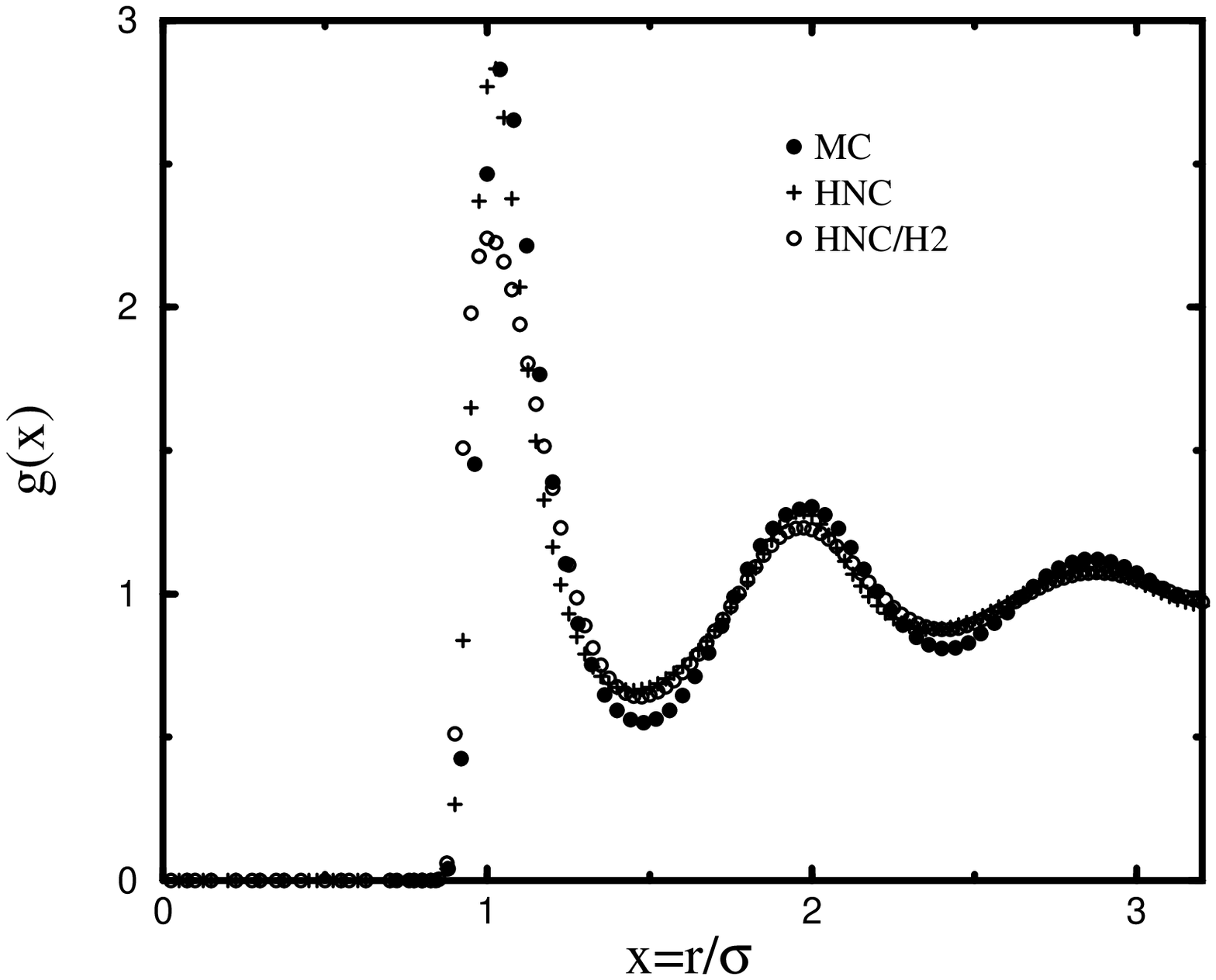}}
\caption[Comparison of the Monte Carlo (MC), the HNC, and HNC/H2 results for 
the pair distribution function of the inverse 12th-power fluid at $z=0.813$.
]{\label{fig:hs-gr-12m}
R. Fantoni and G. Pastore}
\end{figure}

In figure \ref{fig:hs-gr-12m} we compare the MC, the HNC, and the 
HNC/H2 results for the pair distribution function.
 
\subsubsection[The inverse 6th power potential]{The inverse 6th power potential}

The freezing point for this fluid is at $z=1.54$. The RY $\alpha$
parameter to achieve thermodynamic consistency at this value of $z$ is
$1.209$. 

In table \ref{tab:6} we compare various thermodynamic quantities
(the excess internal energy per particle, the excess virial pressure,
the bulk moduli) obtained from the RY, the HNC, and the HNC/H2 closures.
In the MC calculation the excess virial pressure
is 38.8 and the bulk modulus 110.1.

\subsubsection[The inverse 4th power potential]{The inverse 4th power potential}

The freezing point for this fluid is at $z=3.92$. The RY $\alpha$
parameter to achieve thermodynamic consistency at this value of $z$ is
$1.794$. 

In table \ref{tab:4} we compare various thermodynamic quantities
(the excess internal energy per particle, the excess virial pressure,
the bulk moduli) obtained from the RY, the HNC, and the HNC/H2 closures.
In the MC calculation the excess virial pressure
is 108.7 and the bulk modulus 156.

\begin{figure}[hbt]
\centerline{\includegraphics[width=10cm]{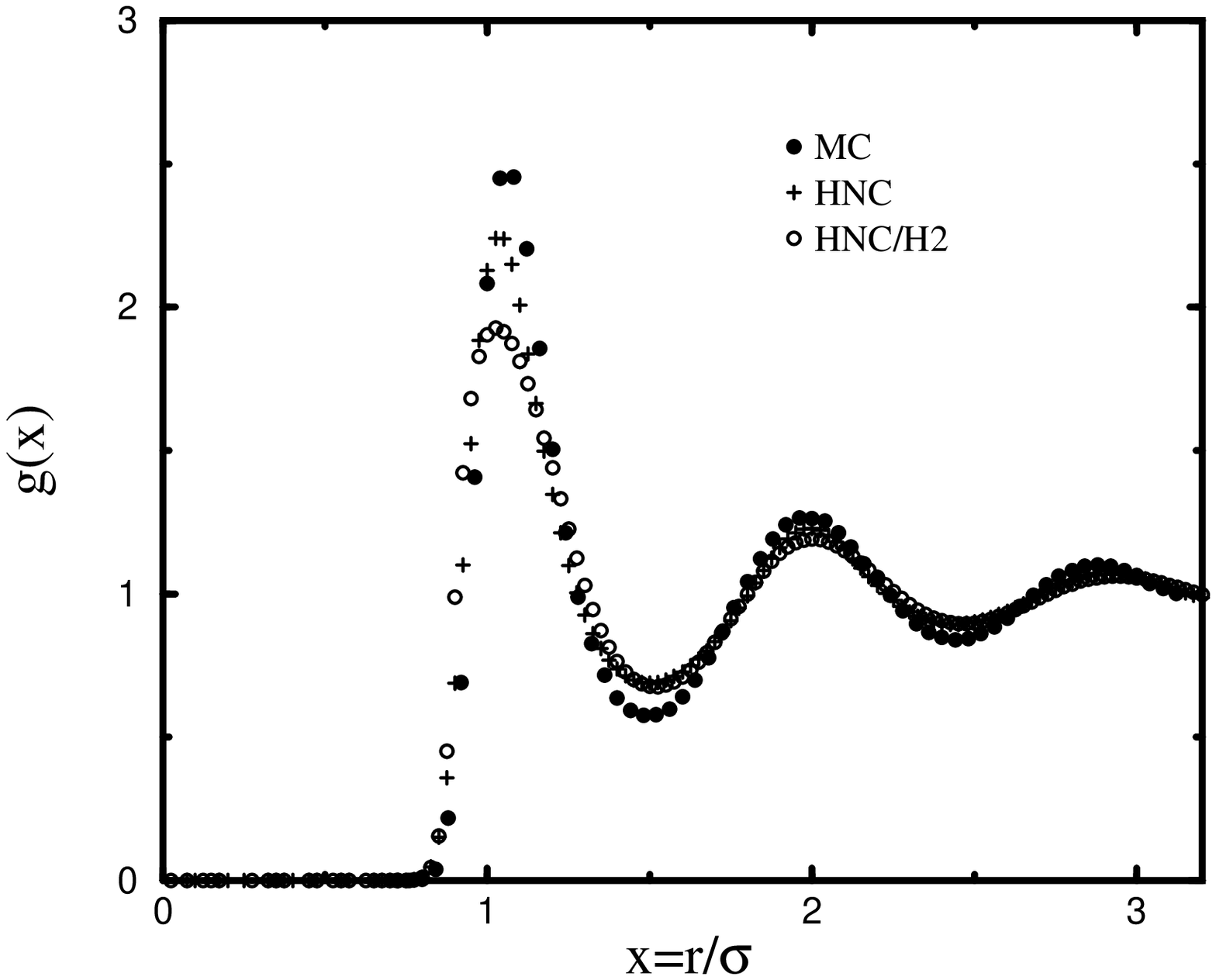}}
\caption[Comparison of the Monte Carlo (MC), HNC, and HNC/H2 results for 
the pair distribution function of the inverse 4th-power fluid at $z=3.92$.
]{\label{fig:hs-gr-4m}
R. Fantoni and G. Pastore}
\end{figure}

In figure \ref{fig:hs-gr-4m} we compare the MC, the HNC, and the
HNC/H2 results for the pair distribution function.

\subsection[The spinodal line]{The spinodal line}

In this subsection we study a pair potential with a minimum 
In particular we chose the Lennard-Jones potential
\bq
\phi(r)=4\epsilon\left[\left(\frac{\sigma}{r}\right)^{12}-\left(
\frac{\sigma}{r}\right)^6\right]~~~,
\eq
where $\epsilon$ and $\sigma$ are positive parameters. The critical 
point for this fluid is at \cite{Potoff98} $T_c^*=1.3120\pm0.0007$,
$\rho_c^*=0.316\pm0.001$, and $P_c^*=0.1279\pm0.0006$.

Integral equations usually fail to have a solution at low temperature and 
intermediate densities, {\sl i.e.} in the two-phases unstable region of the 
phase diagram. In particular it is well known that the HNC approximation 
is unable to reproduce the {\sl spinodal line}, the locus of points of 
infinite compressibility in the phase diagram \cite{Belloni93}. 
This is due to the loss of 
solution as one approaches the unstable region on an isotherm from high 
or from low densities. The line of loss of solution, in the phase diagram,
is called {\sl termination line}. According to the discussion of
section \ref{sec:uniqueness}, the loss of solution for the HNC
approximation can be traced back to the loss of strict convexity of
the generating 
functional \cite{Ferreira94}. Indeed, using HNC approximation, we
computed the bulk modulus from the compressibility equation $B_c$, on
several isotherms as a function of the density. At low temperatures we
found that both at high density and at low density we were unable to
continue the isotherm at low values of $B_c$. Zerah' s algorithm either 
could not get to convergence or it would converge at a non physical 
solution (with a pole in the structure factor at some finite wavevector $k$). 
Since HNC/H2 has, by construction, an always strictly convex
generating functional, we expect it to be able to approximate a
spinodal line (there should be no termination line). 

\begin{figure}[hbt]
\centerline{\includegraphics[width=11cm]{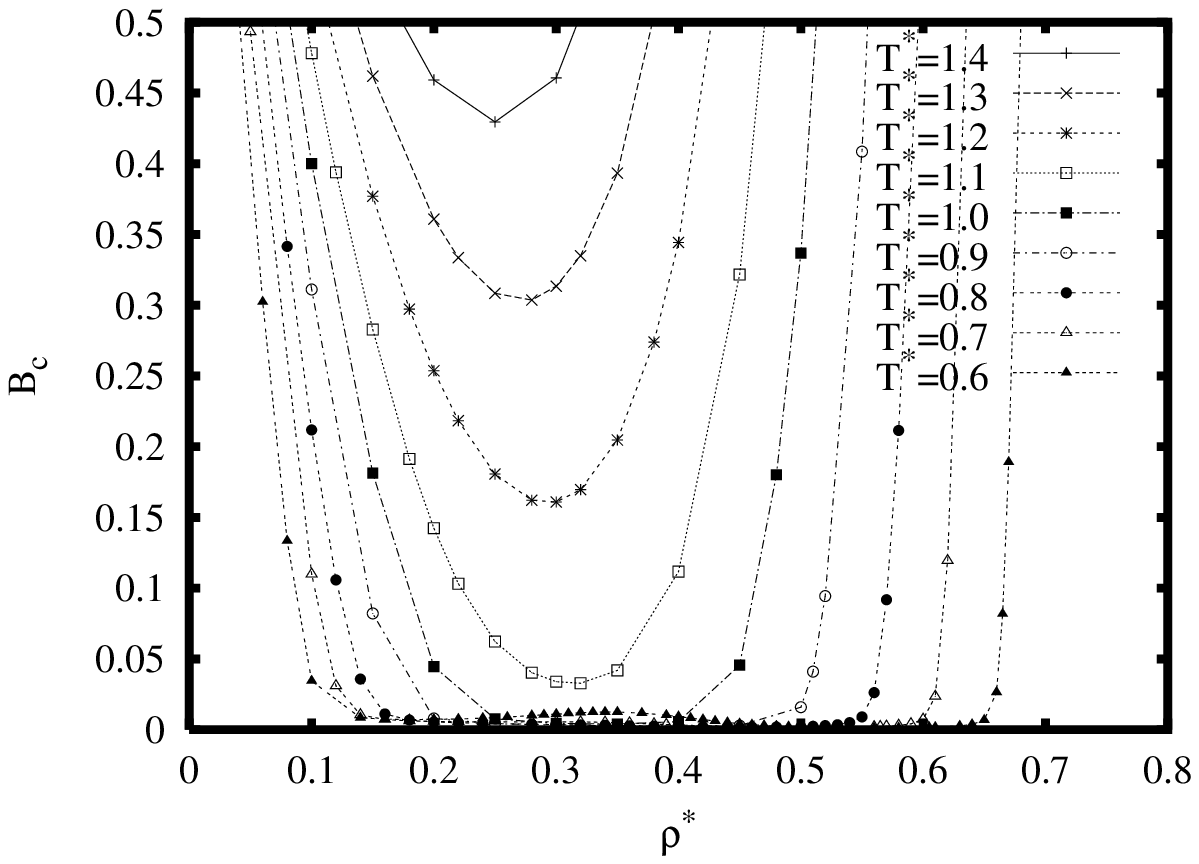}}
\caption[Behavior of $B_c$ of the Lennard-Jones fluid, on several isotherms as a function of the
density for the HNC/H2 approximation.
]{\label{fig:MHNC-Bc1024}
R. Fantoni and G. Pastore}
\end{figure}

In Figure \ref{fig:MHNC-Bc1024} we show the behavior of $B_c$ on several
isotherms as a function of density, calculated with the HNC/H2
approximation. We see that now there are no termination points. $B_c$
never becomes exactly zero and the low temperature isotherms develop a
bump in the intermediate density region. The same plot for the bulk
modulus calculated from the virial pressure $B_p$, shows that at low
temperatures this bulk modulus indeed becomes zero along the isotherms
both at high and low densities.

\begin{figure}[hbt]
\centerline{\includegraphics[width=11cm]{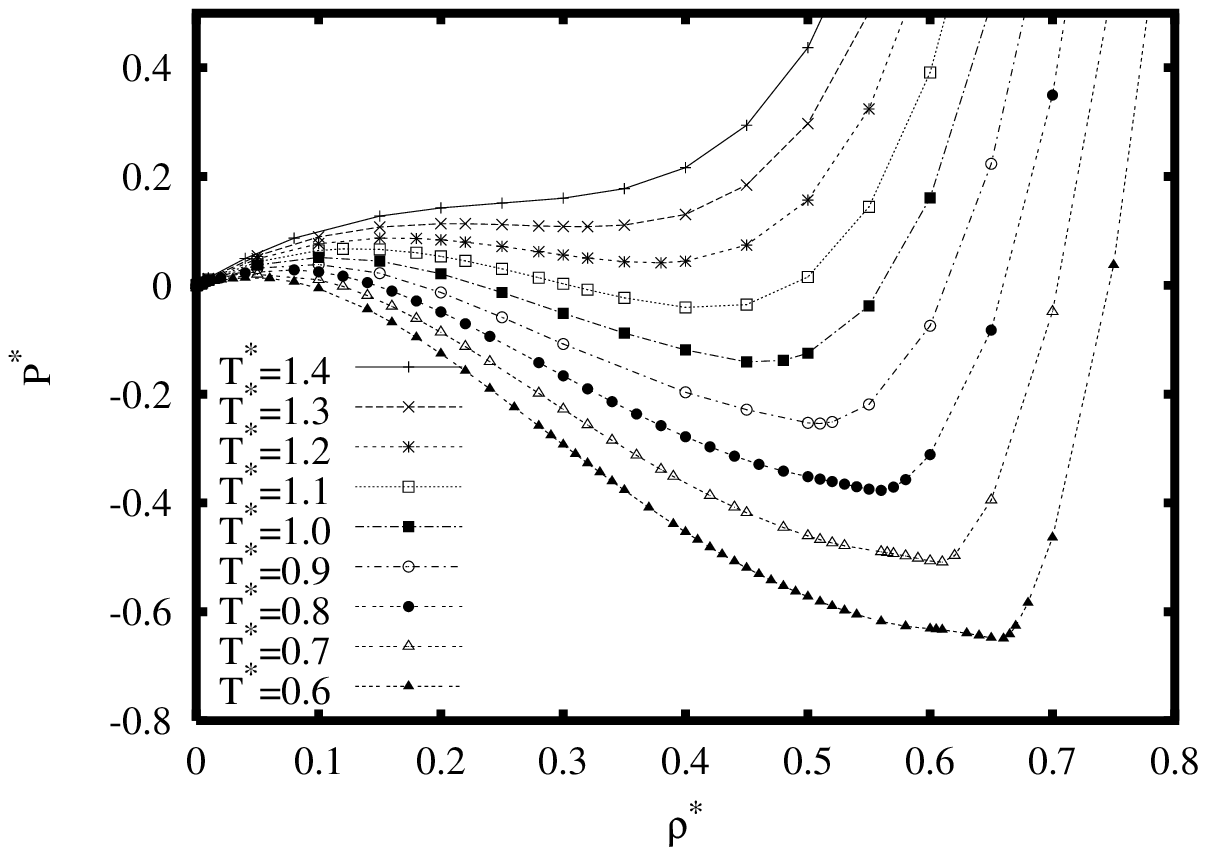}}
\caption[Behavior of the pressure of the Lennard-Jones fluid, on several isotherms as a function of 
the density for the HNC/H2 approximation.
]{\label{fig:MHNC-P1024}
R. Fantoni and G. Pastore}
\end{figure}

In figure \ref{fig:MHNC-P1024} the pressure is plotted as a function of the 
density on several isotherms for the HNC/H2 approximation. Apart from
the fact that we find negative pressures, the isotherms have a van der
Waals like behavior.
 
The graphical analysis of the pressure plotted as a function
of the chemical potential shows that the coexistence of the two phases
(points where the curve crosses itself) is possible and is lost
between $T^*=1.1$ and $T^*=1.2$. There generally are two points of
coexistence. 

\section[Improving the closures]{Improving the closures}
\label{sec:improving}

The numerical results for HNC/H2 exhibit interesting features as far as the
coexistence region is concerned but show unambiguously a worst agreement with
the MC structural data in correspondence with a marginal improvement in the
thermodynamics.

We feel that the main problem is  the difficulty of an accurate 
description of the bridge functions in terms of powers of the pair correlation
function. Recent investigations on improved closures
seem to  point to the indirect correlation function $\gamma(r)$ or some
renormalized version of it, as the best
starting point  for progress. However, before moving to more complex relations 
or functional dependences, we have explored two possible directions for
improving the HNC/H2 closure. In the first approach we have tried to follow the
MHNC approach by Lado {\sl et al.} \cite{Lado73}. In the second we
have explored the 
possibilities of optimization offered by the numerical coefficient of the cubic
term in the generating functional.

\subsection[Pseudo bridge functions for HNC/H2]{Pseudo-bridge functions for
HNC/H2}
\label{sec:reference}

From the graphical analysis of the pair distribution function it is
known \cite{Hansen} that $g(r)$ may be written as
\bq
g(r)=\exp(-\beta\phi(r)+\gamma(r)+B(r))~~~,
\eq
where $\gamma(r)=h(r)-c(r)$ is the sum of all the series type diagrams 
and $B(r)$ the sum of bridge type diagrams. If we take 
\bq
B(r)=-\frac{1}{2}h^2(r)+G(r)~~~,
\eq
we have that our HNC/H2 approximation amounts to setting $G(r)=0$.
Rosenfeld and Ashcroft \cite{Rosenfeld79} proposed that $B(r)$ 
should be essentially the same for all potentials $\phi(r)$. We 
now make a similar proposal for the $G$ function and we will refer to it as
{\em pseudo bridge function}. In the same spirit
of the RHNC approximation of Lado \cite{Lado73} we will approximate
$G(r)$ with the $G$ function of a short range (reference) potential
$\phi_0(r)$. Assuming known the properties of the reference system,
we can calculate the $G$ function as follows
\bq
G_0(r)=\ln\left[g_0(r)e^{\beta\phi_0(r)}\right]-\gamma_0(r)+
\frac{1}{2}h_0^2(r)~~~.
\eq 
The reference HNC/H2 (RHNC/H2) approximation is then
\bq\label{RHNC/H2}
g(r)=\exp(-\beta\phi(r)+\gamma(r)-\frac{1}{2}h^2(r)+G_0(r))~~~.
\eq

An expression for the free energy functional can be obtained 
{\sl turning on} the potential $\phi(r)$ in two stages: first, from 
the noninteracting state to the reference potential $\phi_0(r)$
and then from there to the full potential $\phi(r)$. To this end
we write
\bq
\phi(r;\lambda_0,\lambda_1)=\lambda_0\phi_0(r)+\lambda_1\Delta
\phi(r)~~~,
\eq
with $\Delta\phi(r)=\phi(r)-\phi_0(r)$. Following the same steps
as in \cite{Lado83} we obtain for the excess free energy per 
particle
\bq
f^{exc}=f_1+f_2+f_3^{(0)}+\Delta f_3~~~
\eq
where the first two terms were already encountered in section 
\ref{sec:uniqueness}
\bq \label{f1}
\beta f_1&=&\frac{1}{2}\rho\int d\rr\left\{1+g(r)\left[\ln\left(g(r)
e^{\beta\phi(r)}\right)-1\right]-h^2(r)/2+h^3(r)/6\right\}~~~,\\ \label{f2}
\beta f_2&=&\frac{1}{2\rho}\int\frac{d\kk}{(2\pi)^3}\{\rho\hat{h}(k)-
\ln[1+\rho\hat{h}(k)]\}~~~.
\eq
The third term is assumed known
\bq
\beta f^{(0)}_3=-\frac{1}{2}\rho\int d\rr\int_0^1d\lambda_0
G(r;\lambda_0,0)\frac{\partial g(r;\lambda_0,0)}{\partial\lambda_0}
=\beta (f^{(0)}-f_1^{(0)}-f_2^{(0)})~~~,
\eq
here $f^{(0)}$ is the excess free energy per particle of the reference
system and $f_1^{(0)},f_2^{(0)}$ are defined as in equations (\ref{f1}),
(\ref{f2}) for the reference potential and its corresponding correlation 
functions. The last term is
\bq
\beta\Delta f_3=-\frac{1}{2}\rho\int d\rr\int_0^1d\lambda_1
G(r;1,\lambda_1)\frac{\partial g(r;1,\lambda_1)}{\partial\lambda_1}~~~.
\eq
According to our proposal, $G$ is insensitive to a change in potential
from $\phi_0$ to $\phi$. We may then approximate this last term as 
follows
\bq
\beta\Delta f_3\approx -\frac{1}{2}\rho\int d\rr G_0(r)[g(r)-g_0(r)]~~~.
\eq

Now that we have the free energy we may consider it as a functional of
both $h(r)$ and $G_0(r)$ and take its variation with respect to these 
functions. We find,
\bq \nonumber
\beta\,\delta f^{exc}&=&\frac{1}{2}\rho\int d\rr\left\{c(r)-h(r)+h^2(r)/2
+\ln\left[g(r)e^{\beta\phi(r)}\right]-G_0(r)\right\}\delta h(r)\\
&&-\frac{1}{2}\rho\int d\rr[g(r)-g_0(r)]\delta G_0(r)~~~.
\eq 
It follows that the free energy is minimized when both the RHNC/H2
closure (equation (\ref{RHNC/H2})) is satisfied and when the following
constraint
\bq\label{constraint}
\int d\rr[g(r)-g_0(r)]\delta G_0(r)=0~~~,
\eq
is fulfilled.

Taking the second functional derivative of $f^{exc}$ with respect to 
$h(r)$ we find that also this free energy is a strictly convex functional 
of the total correlation function. This property was lacking in the 
RHNC theory and constitutes the main feature of the RHNC/H2 closure.
As already stressed in section \ref{subsec:HNC/H2} it ensures that 
if a solution to the integral equation exists it has to be unique.

The constraint, as for RHNC, gives a certain thermodynamic consistency
to the theory (see \cite{Lado83}). If we choose a hard sphere 
reference potential
$\phi_0(r)=\phi_0(r;\sigma)$ which depends on the  length scale 
$\sigma$, the optimum values of
the parameters that makes the generating  functional a  free energy 
can be determined by the 
constraint (\ref{constraint}) which becomes
\bq
\int d\rr[g(r)-g_0(r)]\frac{\partial G_0(r)}{\partial\sigma}=0~~~,
\eq  

However, neither the hard-sphere pseudo bridge functions nor  some empirical
attempt to model the unknown function via a Yukawa function  
provided useful results.

\subsection[Optimized HNC/H3 approximation]{Optimized HNC/H3 approximation}

For $\gamma=0$ HNC/H3 reduces to HNC/H2. For $\gamma>0$ the first peak
of the pair distribution function is dumped respect to the one of the
pair distribution function calculated with HNC/H2. For $\gamma<0$ the
first peak increases giving in general a better fit to the simulation
data. 

\begin{figure}[hbt]
\centerline{\includegraphics[width=10cm]{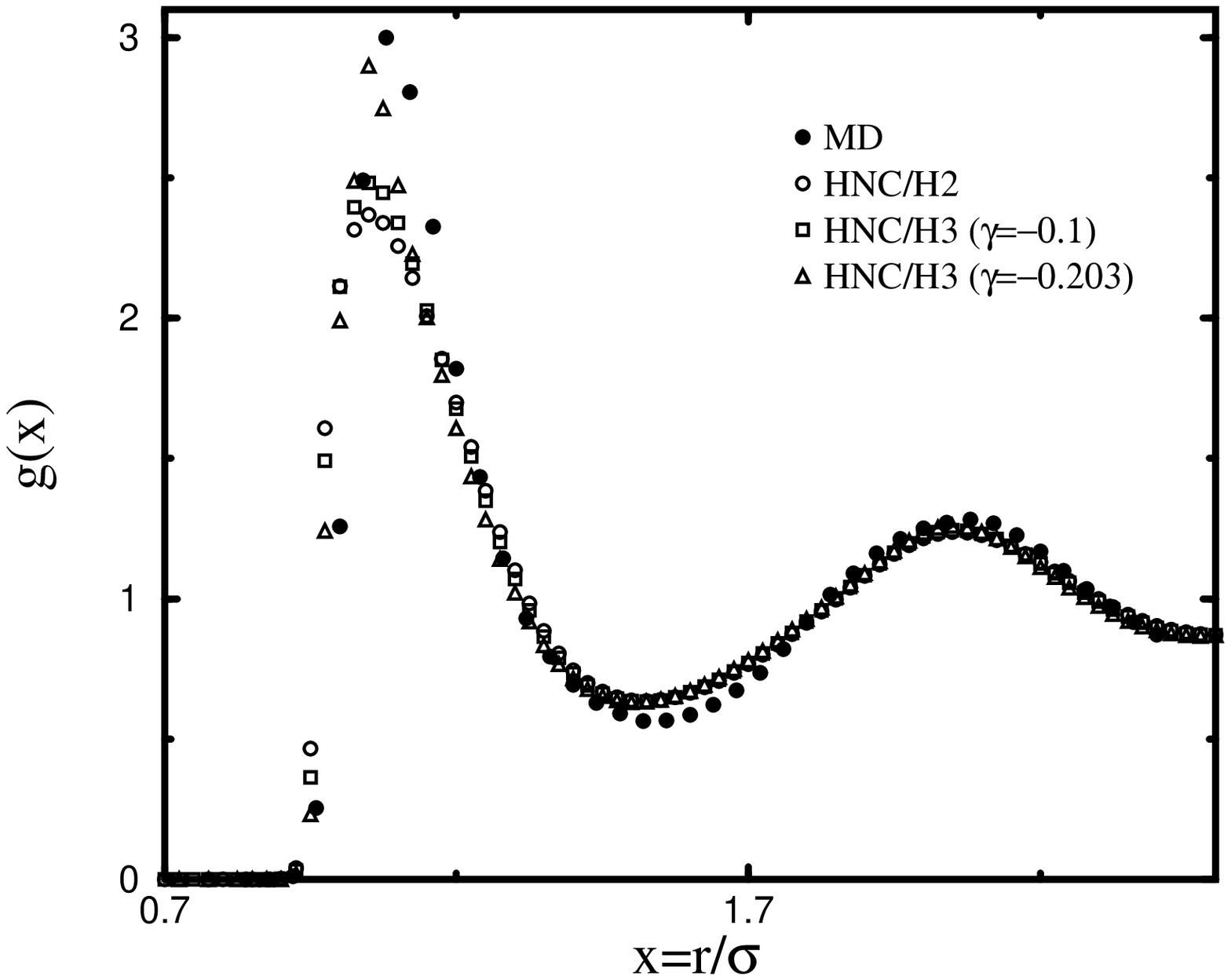}}
\caption[Comparison of the pair distribution function of a Lennard-Jones 
fluid at $\rho^*=0.85$ and $T^*=0.719$, computed from the 
molecular dynamic (MD) simulation of Verlet, the HNC/H2 approximation,
and the HNC/H3 approximation. For HNC/H3 we present results
obtained setting $\gamma=-0.1$ (when the generating functional of the
approximation is still strictly convex) and $\gamma=-0.203$ (which
gives the best fit possible to the simulation data but does not ensure
the strict convexity of the generating functional).
]{\label{fig:HNC/H3}
R. Fantoni and G. Pastore}
\end{figure}

In figure \ref{fig:HNC/H3} we compare the pair distribution
function of the Lennard-Jones fluid near its triple point, calculated
with a molecular dynamic simulation \cite{Verlet68}, the HNC/H2
approximation, the approximation HNC/H3 with $\gamma=-0.203$
(at lower values of $\gamma$ Zerah' s algorithm would fail to
converge), and the approximation HNC/H3 with $\gamma=-0.1$ (when the
generating functional of HNC/H3 is still strictly convex). 
As we can see HNC/H3 fits the simulation data better than
HNC/H2 even if  the first peak is still slightly
displaced to the left of the simulation data, a well known problem of the HNC
approximation \cite{Rosenfeld79}. 

The best results are  given by HNC/H3 with $\gamma=-0.203$. Note that the
HNC/H3 generating functional at this value of $\gamma$ is not
strictly convex (strict convexity is lost for $\gamma\lesssim
-1/9$). The first peak of the static structure factor is at $k\sigma \simeq 
6.75$ and has a magnitude of $2.41$, a quite low value for a liquid near the
triple point. We have calculated
the pressure and the internal energy. We found $\beta P/\rho\simeq
3.87$ and $U^{exc}/(N\epsilon)\simeq -5.72$ (very close to the HNC
results $\beta P/\rho\simeq 3.12$ and $U^{exc}/(N\epsilon)\simeq
-5.87$) to be compared with the simulation results \cite{Verlet67}
$0.36$ and $-6.12$ respectively. The bulk moduli are 
$B_c\simeq 11.74$ and $B_p\simeq 36.61$ which shows that at the
chosen value of $\gamma$ we do not have the thermodynamic consistency
virial-compressibility and we do not improve on HNC inconsistency 
(using HNC we find $B_c\simeq 7.09$ and
$B_p\simeq 32.72$). 

\section[Conclusions]{Conclusions}

In this paper we have analyzed the relations between generating functionals,
thermodynamic consistency and uniqueness of the solution of the integral
equations of liquid state theory. We think that the requirement of deriving from
a free energy and the uniqueness of the solution are two important ingredients
to enforce in the quest for better closures. The former requirement is of course
crucial to get virial-energy consistency. But it is also important to get
integral equations able to provide a closed formula for the chemical potential
without additional approximations. This last issue looks highly desirable for
applications of IET to the determination of phase diagrams.
The latter is certainly a useful constraint from the numerical point of view but
it is also a very strong condition, probably able to avoid some non  physical
behavior in the coexistence region, although this point 
would deserve further investigation.
 
In this work, we have started an exploration of the capabilities of the combined
requirement of consistency and uniqueness, starting  with simple modifications 
to the HNC closure, 
corresponding to the addition of a square and a cubic power of 
$h(r)$ in 
the HNC functional. 
We found a couple of approximations (HNC/H$2$ and HNC/H$3$),
which have built in the virial-free energy thermodynamic consistency  and
have a unique solution.

We numerically tested these closures on inverse power and the 
Lennard-Jones  fluid.
From the tests on the inverse power potential fluids one can
see that the HNC/H2 approximation is comparable to HNC for the thermodynamic
quantities and performs worst than RY and even  
HNC for structural properties. The tests on the Lennard-Jones
fluid revealed as this approximation does not suffer from the presence
of a termination line (present in HNC and almost all the existing closures). 
This
allowed us to follow isotherms from the low density  to the high
density region and this behavior would be very useful in the  study of the 
phase coexistence. 
However, the thermodynamic results show only a marginal improvement on HNC and
the structure is definitely worse.

Our
trials to improve HNC/H2 in the same spirit of the modified HNC approaches  
did not succeed. We feel that the main reason is in the difficulty of 
modeling the
real bridge functions through a polynomial in the function $h(r)$. In this
respect, approaches based on generating functionals depending on the indirect
correlation function $\gamma(r)$ look more promising but we have not tried them
yet.

Much better results for the structure are found with HNC/H3 as is shown
in figure \ref{fig:HNC/H3}. However, probably for the same reasons just
discussed, one has to renounce to have an approximation with a strictly
convex generating functional depending on $h(r)$. 
The thermodynamics reproduced by HNC/H3 is not yet satisfactory: 
due to the slight left shift of
the main peak of the $g(r)$ the calculated pressure misses the
simulation result. Nonetheless the presence of the
free parameter $\gamma$ in HNC/H3 leaves open the possibility of
imposing the thermodynamic consistency virial-compressibility. If the
value of the parameter needed to have the consistency is bigger than
$-1/[3\sup g(r)]$ then we would have an approximation which is
completely thermodynamically consistent and have a unique solution.
This strategy may eventually lead to discover that the price we have to
pay to have a completely thermodynamically consistent approximation is
the loss of strict convexity of the generating functional.

\appendix
\renewcommand{\theequation}{\Alph{section}.\arabic{equation}}

\section[Appendix: Generating functionals of $\gamma$]{Appendix: Generating functionals of $\gamma$}
\label{app:gammafunct}

Often in the numerical solution of the OZ plus closure integral equation
use is made of the auxiliary function $\gamma(r)=h(r)-c(r)$. Suppose 
that the closure relation can be written as 
\bq
\rho^2 c(r)=-\Psi\{\gamma(r)\}~~~,
\eq
where $\Psi$ is a local function of the  function $\gamma$ and has a dependence
on the value of the pair potential not explicitly shown. 

We want to translate the integral equation into a variational principle
involving functionals of $\gamma(r)$. Then we introduce a closure 
functional $\ff_{cl}[\gamma]$ such that
\bq
\frac{\delta \ff_{cl}[\gamma]}{\delta \gamma(r)}=\Psi\{\gamma(r)\}~~~,
\eq
and an OZ functional $\ff_{OZ,c}[\gamma]$ such that, when $c(r)$ and 
$\gamma(r)$ satisfy the OZ equation, we have 
\bq
\frac{\delta \ff_{OZ,c}[\gamma]}{\delta \gamma(r)}=\rho^2 c(r)~~~.
\eq
Then when both the closure and the OZ relations are satisfied, the 
functional $\ff=\ff_{cl}+\ff_{OZ,c}$ is stationary with respect to 
variations of $\gamma(r)$, {\sl i.e.}
\bq
\frac{\delta \ff[\gamma]}{\delta \gamma(r)}=0~~~.
\eq
This is the variational principle sought.

Now, we want to find $\ff_{OZ,c}$. The OZ equation in $k$ space is
\bq
\rho \hat{c}^2(k)+\rho\hat{\gamma}(k)\hat{c}(k)-\hat{\gamma}(k)=0~~~.
\eq
When we solve it for $\hat{c}$ we find two solutions
\bq
\hat{c}=\frac{-\hat{\Gamma}\pm\sqrt{\hat{\Gamma}^2+4\hat{\Gamma}}}
{2\rho}~~~,
\eq
where $\hat{\Gamma}(k)=\rho\hat{\gamma}(k)$ is always positive 
since
\bq
\hat{\Gamma}=\rho^2\hat{h}\hat{c}=\rho^2\frac{\hat{h}^2}{1+\rho\hat{h}}
=\rho^2\frac{\hat{h}^2}{S(k)}~~~,
\eq
$S(k)$ being the liquid static structure factor which is positive 
definite for all $k$.
Since $\hat{c}(k)$ is a function which oscillates around 0, where 
$\hat{c}$ is negative we have to choose the solution with the minus
sign, where it is positive the one with the plus sign. In particular,
if the isothermal compressibility of the liquid $\chi_T$, is smaller 
than the one of the ideal gas $\chi_T^0$, we have that 
\bq
\hat{c}(0)=\frac{1}{\rho}\left(1-\frac{\chi_T^0}{\chi_T}\right)<0~~~,
\eq
and we have to start with the minus sign.

The functional we are looking for is then (see equation
(30) in \cite{Olivares76}, with the constant set equal to zero)
\bq
\ff_{OZ,c}[\gamma]=\int_0^1dt\int d\rr\,\gamma(r)\int 
\frac{d\kk}{(2\pi)^3}\frac{\rho}{2}e^{i\kk\cdot\rr}\left[-t\hat{\Gamma}(k)
+s_c(k)\sqrt{t^2\hat{\Gamma}^2(k)+4t\hat{\Gamma}(k)}\right]~~~,
\eq
where $s_c(k)$ is $+1$ when $\hat{c}(k)\ge 0$ and $-1$ when $\hat{c}(k)<0$. 
Rearranging the integrals and making the change of variable 
$y=t\hat{\Gamma}$ we find
\bq \nonumber
\ff_{OZ,c}[\gamma]&=&\frac{1}{2}\int\frac{d\kk}{(2\pi)^3}
\int_0^{\hat{\Gamma}(k)}dy\left(-y+s_c(k)\sqrt{y^2+4y}\right)\\ \nonumber
&=&\int\frac{d\kk}{(2\pi)^3}\left\{-\hat{\Gamma}^2/4+s_c(k)
\left[\left(1+\hat{\Gamma}/2\right)\sqrt{\left(1+
\hat{\Gamma}/2\right)^2-1}\right.\right.\\
&&\hspace{1.8cm}\left.\left.-\ln\left(1+\hat{\Gamma}/2+ \sqrt{\left(1+
\hat{\Gamma}/2\right)^2-1}\right)\right]\right\}~~~. \label{fozc}
\eq

If the closure relation has the form
\bq
\rho^2 h(r)=-\Psi\{\gamma(r)\}~~~,
\eq
we can derive the corresponding functional using the same procedure. 
The final result is a functional $\ff_{OZ,h}[\gamma]$ which differs from
(\ref{fozc}) for a plus sign in front of the first term in the integral. 

However, by examining their second functional derivatives, 
we notice that both $\ff_{OZ,c}[\gamma]$ and $\ff_{OZ,h}[\gamma]$
are not certainly convex or concave. Thus, any check of the convexity properties
of generating functionals of the $\gamma(r)$ function should be done on the full
functional.

\section[Appendix: Thermodynamic consistency]{Appendix: Thermodynamic consistency}
\label{app:consistency}

For a homogeneous liquid interacting through a pair potential $\phi(r)$,
the Helmholtz free energy per particle $f$ can be considered a functional 
of $\phi$. Indeed, in the canonical ensemble, one has
\bq
\beta f[\phi]=\beta f_0-\frac{1}{N}\ln\left(\frac{1}{V^N}
\int \exp\left[-\beta\frac{1}{2}\sum_{i\neq j}\phi(\rr_{ij})\right]\,
d\rr_1\cdots d\rr_N\right)~~~,
\eq
where $f_0$ is the free energy per particle of the ideal gas 
$(\phi=0)$ and $V$ is the volume of the liquid. Taking the functional
derivative with respect to $\beta\phi(r)$ one finds
\bq
\frac{\delta\beta f[\phi]}{\delta\beta \phi(r)}=\frac{\rho}{2}g(r)~~~.
\eq

Imagine that we found a functional ${\cal A}([h],[\phi],\rho,\beta)$
that has an extremum for those correlation functions that solve the OZ
and the closure system of equations. Suppose further that such
functional has the following property
\bq
\frac{\delta\beta {\cal A}}{\delta\beta \phi(r)}=\frac{\rho}{2}g(r)~~~,
\eq
which can be rewritten more explicitly as follows
\bq
\left.\frac{\delta\beta{\cal A}}{\delta\beta\phi(r)}\right|_{[h],\rho,\beta}
+\int d\rr^\prime\left.\frac{\delta\beta{\cal A}}{\delta h(r^\prime)}
\right|_{[\phi],\rho,\beta}\frac{\delta h(r^\prime)}{\delta\beta\phi(r)}
=\frac{\rho}{2}g(r)~~~.
\eq
Evaluating this expression on the correlation function $\bar{h}$
solution of the OZ plus closure system of equations, which is an
extremum for ${\cal A}$, we find
\bq \label{app:dadphi}
\left.\frac{\delta\beta{\cal A}}{\delta\beta\phi(r)}\right|
_{[\bar{h}],\rho,\beta}=\frac{\rho}{2}\bar{g}(r)~~~.
\eq
Then we can write
\bq
\beta {\cal A}([\bar{h}],[\phi],\rho,\beta)=\int 
d\rr\left.\frac{\delta\beta {\cal A}}{\delta\beta \phi(r)}\right|
_{[\bar{h}],\rho,\beta}
\beta\phi(r)+{\cal D}([\bar{h}],\rho,\beta)~~~,
\eq
with ${\cal D}$ a functional independent of $\phi$. Changing variables
to adimensional ones, $\rr=\rr^*\rho^{-1/3}$ and using equation
(\ref{app:dadphi}) we find
\bq
\beta {\cal A}([\bar{h}^*],[\phi],\rho,\beta)=\frac{1}{2}\int d\rr^*\,
\bar{g}^*(r^*)\beta\phi(r^*\rho^{-1/3})+
{\cal D}([\bar{h}^*],\rho,\beta)~~~,
\eq
where we defined new distribution functions $g^*(r^*)=
g(r^*\rho^{-1/3})$. If ${\cal D}$ has no explicit dependence on $\rho$
then one readily finds
\bq\nonumber
\rho\frac{\partial\beta{\cal A}([\bar{h}^*],[\phi],\rho,\beta)}
{\partial\rho}
&=&-\frac{\rho}{6}\int d\rr^*\,\bar{g}^*(r^*)\beta
\phi^\prime(r^*\rho^{-1/3})r^*\rho^{-4/3}\\\nonumber
&=&-\frac{\rho}{6}\int d\rr\,\bar{g}(r)\,\beta\phi^\prime(r)\,r\\
\label{app1P}
&=&\beta P^{exc}/\rho~~~,
\eq
where again we used the fact that ${\cal A}$ has an extremum for
$h=\bar{h}$. We used a prime to denote a derivative with respect 
to the argument and $P^{exc}$ is the excess pressure of the liquid.

If ${\cal D}$ has no explicit dependence on $\beta$ we also find
\bq\nonumber
\frac{\partial \beta{\cal A}([\bar{h}^*],[\phi],\rho,\beta)}
{\partial\beta}
&=&\frac{\rho}{2}\int d\rr\,\bar{g}(r)\phi(r)\\ \label{app1e}
&=&U^{exc}/N~~~,
\eq
where $U^{exc}$ is the excess internal energy.

If ${\cal D}$ has no explicit dependence on both $\beta$ and $\rho$,
${\cal D}([\bar{h}^*],\rho,\beta)={\cal D}([\bar{h}^*])$, we
conclude from equations (\ref{app1P}) and (\ref{app1e}) that 
\bq
{\cal A}([\bar{h}^*],[\phi],\rho,\beta)=f^{exc}(\rho,\beta)+
{\rm constant}
~~~,
\eq
where $f^{exc}$ is the excess free energy per particle of the fluid.
Under these circumstances we see from equation (\ref{app1P}) that we
have thermodynamic consistency between the route to the pressure going
through the partial derivative of the free energy and the route to
the pressure going through the virial theorem.

\section{Appendix: Strict convexity of $\ff_{OZ}[h]$}
\label{app:convexity}

It can be proven that the functional
\bq \label{app2FOZ}
\ff_{OZ}[h]=\int\frac{d\kk}{(2\pi)^3}\{\rho\hat{h}(k)-
\ln[1+\rho\hat{h}(k)]\}~~~,
\eq
defined on the convex set
\bq
D_c=\{h(r)|S(k)>0~~~\forall k\}~~~,
\eq
is a strictly convex functional. The strict convexity is a trivial
consequence of the strict convexity of the integrand in equation
(\ref{app2FOZ}).

It remains to prove that $D_c$ is a convex set. Given two elements
of this set $h^\prime$ and $h^{\prime\prime}$, we need to show
that $h=\lambda h^\prime+(1-\lambda)h^{\prime\prime}$ is an element
of $D_c$ for all $\lambda\in[0,1]$. Since
\bq\nonumber
S(k)&=&1+\rho\hat{h}(k)\\\nonumber
&=&1+\rho[\lambda\hat{h}^\prime(k)+(1-\lambda)\hat{h}^{\prime\prime}(k)]
\\\nonumber
&=&1+\lambda[S^\prime(k)-1]+(1-\lambda)[S^{\prime\prime}(k)-1]\\
&=&\lambda S^\prime(k)+(1-\lambda)S^{\prime\prime}(k) >0
~~~\forall \lambda\in[0,1]~~~,
\eq
then $D_c$ is a convex set.

\bibliography{ietvar}

\begin{thebibliography}{10}

\bibitem{Hansen}
J.~P. Hansen and I.~R. McDonald.
\newblock {\em {"Theory of simple liquids"}}.
\newblock Academic Press, London, 2nd edition, 1986.

\bibitem{Caccamo}
C.~Caccamo.
\newblock {\em Phys. Rep.}, {\bf 274}:1, (1996).

\bibitem{BornGreen}
{M. Born and H. S. Green}.
\newblock {\em Proc. Roy. Soc.}, {\bf A188}:10, (1946).

\bibitem{PY}
{J. K. Percus and G. J. Yevick}.
\newblock {\em Phys. Rev.}, {\bf 110}:1, (1958).

\bibitem{HNC}
{J. M. J. van Leeuwen, J. Groenveld, and J. De Boer}.
\newblock {\em Physica}, {\bf 25}:792, (1959).

\bibitem{Rosenfeld79}
Y.~Rosenfeld and N.~W. Ashcroft.
\newblock {\em Phys. Rev. A}, {\bf 20}:1208, (1979).

\bibitem{Lado}
{F. Lado, S. M. Foiles, and N. W. Ashcroft}.
\newblock {\em Phys. Rev.}, {\bf A28}:2374, (1983).

\bibitem{Verlet}
L.~Verlet.
\newblock {\em Mol. Phys.}, {\bf 41}:183, (1980).

\bibitem{MS}
{G. A. Martynov and G. N. Sarkisov}.
\newblock {\em Mol. Phys.}, {\bf 49}:1495, (1983).

\bibitem{Duh}
{D. -M. Duh and A. D. J. Haymet}.
\newblock {\em J. Chem. Phys.}, {\bf 103}:2625, (1995).

\bibitem{Morita61}
{T. Morita and K. Hiroike}.
\newblock {\em Prog. Theor. Phys.}, {\bf 25}:537, (1961).

\bibitem{Belloni93}
L.~Belloni.
\newblock {\em J. Chem. Phys.}, {\bf 98}:8080, (1993).

\bibitem{AshcroftPoll}
{P. D. Poll and N. W. Ashcroft}.
\newblock {\em Phys. Rev.}, {\bf 35}:5167, (1987).

\bibitem{TDG}
{A. Schlijper, M. Telo de Gama, and P. Ferreira}.
\newblock {\em J. Chem. Phys.}, {\bf 98}:1534, (1989).

\bibitem{Olivares76}
W.~Olivares and D.~A. McQuarrie.
\newblock {\em J. Chem. Phys.}, {\bf 65}:3604, (1976).

\bibitem{Hoye}
{S. H{\o}ye and G. Stell}.
\newblock {\em J. Chem. Phys.}, {\bf 67}:439, (1977).

\bibitem{MSA}
G.~Pastore.
\newblock {\em Mol. Phys.}, {\bf 63}:731, (1988).

\bibitem{ORPA}
{G. Pastore, O. Akinlade, F. Matthews, and Z. Badirkhan}.
\newblock {\em Phys. Rev. E}, {\bf 57}:460, (1998).

\bibitem{Kjellander}
{R. Kjellander, S. Sarman}.
\newblock {\em J. Chem. Phys.}, {\bf 90}:2768, (1989).

\bibitem{Lee92}
{L. Lee}.
\newblock {\em J. Chem. Phys.}, {\bf 97}:8606, (1992).

\bibitem{Verlet68}
L.~Verlet.
\newblock {\em Phys. Rev.}, {\bf 165}:201, (1968).

\bibitem{footnote}
{Our first trial should really be $c=-\ln g-\beta\phi$. Which should be called
  HNC/H1. We have tested numerically this closure and we found that it
  performed worst than HNC/H2 both for the structure and for the thermodynamics
  of the system under exam.}

\bibitem{Zerah85}
G.~Zerah.
\newblock {\em J. Comp. Phys.}, {\bf 61}:280, (1985).

\bibitem{Lado71}
F.~Lado.
\newblock {\em J. Comp. Phys.}, {\bf 8}:417, (1971).

\bibitem{Rogers84}
F.~J. Rogers and D.~A. Young.
\newblock {\em Phys. Rev. A}, {\bf 30}:999, (1984).

\bibitem{Hansen73}
J.~P. Hansen and D.~Shiff.
\newblock {\em Mol. Phys.}, {\bf 25}:1281, (1973).

\bibitem{Potoff98}
J.~J. Potoff and A.~Z. Panagiotopoulos.
\newblock {\em J. Chem. Phys.}, {\bf 109}:10914, (1998).

\bibitem{Ferreira94}
{P. G. Ferreira, R. L. Carvalho, M. M. Telo de Gama, and A. G. Schlijper}.
\newblock {\em J. Chem. Phys.}, {\bf 101}:594, (1994).

\bibitem{Lado73}
F.~Lado.
\newblock {\em Phys. Rev. A}, {\bf 8}:2548, (1973).

\bibitem{Lado83}
{F. Lado, S. M. Foiles, and N. W. Ashcroft}.
\newblock {\em Phys. Rev. A}, {\bf 28}:2374, (1983).

\bibitem{Verlet67}
L.~Verlet.
\newblock {\em Phys. Rev.}, {\bf 159}:98, (1967).

\end{thebibliography}
\bibliographystyle{unsrt}
\newpage

\listoftables

\newpage
\begin{table}[hbt]
\centerline{
\begin{tabular}{|l|l|l|l|l|} \hline\hline
\multicolumn{1}{|c|}{closure}&
\multicolumn{1}{c|}{$U^{exc}/(N\epsilon)$}&
\multicolumn{1}{c|}{$\beta P^{(v)}/\rho-1$}&
\multicolumn{1}{c|}{$B_c$}&
\multicolumn{1}{|c|}{$B_p$}\\ \hline
RY $(\alpha=0.603)$& 2.626 & 18.359 & 69.782 & 70.125\\ \hline
HNC  & 3.009 & 21.036 & 45.278 & 80.430\\ \hline
HNC/H2 & 3.200 & 22.372 & 52.661 & 87.255\\ \hline \hline
\end{tabular}
}
\caption[We compare various thermodynamic quantities as obtained 
from the RY, the HNC, and the HNC/H2 closure, for the inverse 12th-power 
fluid at the freezing point $(z=0.813)$. $U^{exc}/(N\epsilon)$ is the
excess internal energy per particle, $\beta P^{(v)}/\rho-1$ the excess
virial pressure, $B_c$ and $B_p$ are the bulk moduli calculated from the
compressibility and the virial equations respectively. 
]{We compare various thermodynamic quantities as obtained 
from the RY, the HNC, and the HNC/H2 closure, for the inverse 12th-power 
fluid at the freezing point $(z=0.813)$. $U^{exc}/(N\epsilon)$ is the
excess internal energy per particle, $\beta P^{(v)}/\rho-1$ the excess
virial pressure, $B_c$ and $B_p$ are the bulk moduli calculated from the
compressibility and the virial equations respectively. 
\label{tab:12}}
\end{table}
\newpage
\begin{table}[hbt]
\centerline{
\begin{tabular}{|l|l|l|l|l|} \hline\hline
\multicolumn{1}{|c|}{closure}&
\multicolumn{1}{c|}{$U^{exc}/(N\epsilon)$}&
\multicolumn{1}{c|}{$\beta P^{(v)}/\rho-1$}&
\multicolumn{1}{c|}{$B_c$}&
\multicolumn{1}{|c|}{$B_p$}\\ \hline
RY $(\alpha=1.209)$& 4.114 & 39.027 & 110.952 & 111.420\\ \hline
HNC  & 4.235 & 40.178 & 84.016 & 113.733\\ \hline
HNC/H2 & 4.283 & 40.635 & 88.289 & 115.757\\ \hline \hline
\end{tabular}
}
\caption[We compare various thermodynamic quantities as obtained 
from the RY, the HNC, and the HNC/H2 closure, for the inverse 6th-power 
fluid at the freezing point $(z=1.54)$. $U^{exc}/(N\epsilon)$ is the
excess internal energy per particle, $\beta P^{(v)}/\rho-1$ the excess
virial pressure, $B_c$ and $B_p$ are the bulk moduli calculated from the
compressibility and the virial equations respectively.
]{We compare various thermodynamic quantities as obtained 
from the RY, the HNC, and the HNC/H2 closure, for the inverse 6th-power 
fluid at the freezing point $(z=1.54)$. $U^{exc}/(N\epsilon)$ is the
excess internal energy per particle, $\beta P^{(v)}/\rho-1$ the excess
virial pressure, $B_c$ and $B_p$ are the bulk moduli calculated from the
compressibility and the virial equations respectively.
\label{tab:6}}
\end{table}
\newpage
\begin{table}[hbt]
\centerline{
\begin{tabular}{|l|l|l|l|l|} \hline\hline
\multicolumn{1}{|c|}{closure}&
\multicolumn{1}{c|}{$U^{exc}/(N\epsilon)$}&
\multicolumn{1}{c|}{$\beta P^{(v)}/\rho-1$}&
\multicolumn{1}{c|}{$B_c$}&
\multicolumn{1}{|c|}{$B_p$}\\ \hline
RY $(\alpha=1.794)$& 8.001 & 104.664 & 250.106 & 242.948 \\ \hline
HNC  & 8.047 & 105.277 & 223.328 & 244.212\\ \hline
HNC/H2 & 8.068 & 105.542 & 226.966 & 257.678\\ \hline \hline
\end{tabular}
}
\caption[We compare various thermodynamic quantities as obtained 
from the RY, the HNC and the HNC/H2 closure, for the inverse 4th-power 
fluid at the freezing point $(z=3.92)$. $U^{exc}/(N\epsilon)$ is the
excess internal energy per particle, $\beta P^{(v)}/\rho-1$ the excess
virial pressure, $B_c$ and $B_p$ are the bulk moduli calculated from the
compressibility and the virial equations respectively.
]{We compare various thermodynamic quantities as obtained 
from the RY, the HNC and the HNC/H2 closure, for the inverse 4th-power 
fluid at the freezing point $(z=3.92)$. $U^{exc}/(N\epsilon)$ is the
excess internal energy per particle, $\beta P^{(v)}/\rho-1$ the excess
virial pressure, $B_c$ and $B_p$ are the bulk moduli calculated from the
compressibility and the virial equations respectively.
\label{tab:4}}
\end{table}
%

\newpage

\listoffigures

\newpage
\newpage
\newpage
\newpage
\newpage
%
\end{document}